**Fractional Power-Law Spectral Response of $CaCu_3Ti_4O_{12}$ Dielectric: Many-Body Effects**

Jitender Kumar and A.M. Awasthi[*]

Thermodynamics Laboratory, UGC-DAE Consortium for Scientific Research
University Campus, Khandwa Road, Indore- 452 001 (India)

## ABSTRACT

Spectral character of dielectric response in $CaCu_3Ti_4O_{12}$ across 0.5Hz-4MHz over 45-200K corresponding to neither the Debyean nor the KWW relaxation patterns rather indicates a random-walk like diffusive dynamics of moments. Non-linear relaxation here is due to the many body dipole-interactions, as confirmed by spectral-fits of our measured permittivity to the Dissado-Hill behaviour. Fractional power-laws observed in $e^*(w)$ macroscopically reflect the fractal microscopic configurations. Below ~100K, the power-law exponent $m$ ($n$) steeply decreases (increases), indicating finite length-scale collective response of moment-bearing entities. At higher temperatures, $m$ gradually approaches 1 and $n$ falls to low values, reflecting tendency towards the single-particle/Debyean relaxation.

The giant dielectric $CaCu_3Ti_4O_{12}$ (CCTO) reported by Subramanian[1] as a non-ferroelectric and non-relaxor material has importance in many applications[2-3] like micro-capacitors and gas sensing devices. CCTO shows a large dielectric constant over a broad frequency-band and temperature-range from 100-600K in both the polycrystalline as well as single crystal specimens.[1-4] In CCTO below ~100K however, there is a huge drop in the dielectric constant without any structural phase transition,[4] being dispersive with the probing frequency. Intensive research has been focusing to understand the origin of huge dielectric constant. Several possible mechanisms are suggested; according to Homes[4] et. al., it may be due to relaxor-like dynamical slowing down of the dipolar fluctuations in nanosized domains, pertaining to the extrinsic effects like electrode polarization[5] because of different work functions of electrodes and sample. From Sinclair[6] group's point of view, the internal barrier layer capacitor (IBLC) created by the formation of semiconducting grains (surrounded by the insulating grain boundaries) is responsible for the high dielectric constant; the latter being significant for applications as well as from basic science points of view. Phenomenological mechanisms for the colossal dielectric constant (CDC) of CCTO and its abrupt fall at low-$T$'s are open for microscopic description. CCTO is highly disordered material at the nano-scale[7] as well, due to the site/anti-site admixture of Ca- and Cu-ions, enhancing which improves the dielectric properties.[7-8] This disorder makes CCTO electrically heterogeneous and a highly nonlinear material, with a huge $a = d\log(I)/d\log(V)$ coefficient ever observed,[3] with vastly superior applicability. The non-linearity here is intrinsic, as semiconducting and insulating intra-grain locations have been resolved in the tunneling microscopy.[9-10] However, the disorder-borne non-linearity has not yet been recognized in the dielectric spectra of CCTO, crucial for its understanding and the relevant-property-control. Conventional Debyean exponential decay as the relaxation mechanism is not applicable here due to the disorder and cooperative motion of the relaxing entities. To understand the intrinsic heterogeneities, we analyze our CCTO dielectric spectra with the Dissado-Hill (D-H) formulation[11] featuring fractional power-laws, applicable to the cooperative response in the disordered materials, characterizing their microscopic interactions.

---

[*] Corresponding Author: amawasthi@csr.res.in

The ceramic CCTO samples were prepared from high purity (99.99%) powders of $CaCO_3$, CuO, and $TiO_2$ by the conventional solid state route. The pelletized samples (10mm diameter and 1-3mm thick) were sintered at 1100 °C and silver-coated for good electrical contacts for the dielectric measurements. XRD of the samples has been done by RIGAKU rotating anode powder diffractometer with Cu-$K_a$ radiation ($\lambda$ =1.54Å). Dielectric measurements over 10K-room temperature were performed using NOVO-CONTROL (Alpha-A) High Performance Frequency Analyzer across 0.5Hz to 2MHz, using 1V ac signal for excitation.

The Debyean relaxation is inapplicable to the disordered materials with interacting dipoles, as the power-law (non-Debyean) behaviour of the dielectric permittivity $\varepsilon^* = \varepsilon' + i\varepsilon''$ is observed in several disordered dielectrics.[12-14] Here, the dielectric loss $\varepsilon''$ in frequency domain can be described by the following power-laws:

$$\varepsilon''(\omega) \approx (\omega/\omega_p)^{n-1}, \quad \omega > \omega_p \qquad (1)$$

$$\varepsilon''(\omega) \approx (\omega/\omega_p)^{m}, \quad \omega < \omega_p \qquad (2)$$

$\omega_p$ is the frequency corresponding to the maximum dielectric-loss rate. The limiting exponents $n = 0$ and $m = 1$ correspond to the Debyean exponential relaxation of the non-interacting (single) dipoles, fairly explaining the behaviour of weak dipolar solutions or of the gaseous-state dipoles. In real solids having interacting dipoles, the situation quite changes the parameters $n$ and $m$. For most of the materials, the parameters are independent of the temperature in their thermodynamically stable phase. The Dissado-Hill (D-H) many-body model[11] contains the parameters in relations (1) and (2), and explains the non-Debyean behavior when interactions are present between the relaxing entities of the system. D-H model concerns the energy transfer from the quickly-excitable single-dipoles to their slowly-reverberating collective degrees of freedom.[11-17] Lower-frequency cascading down of the energy rather spontaneously gained by the individual dipoles from the applied field is determined by the disorder, via cooperative dynamic structure-modulations. The incipient energy-transformation (from potential to kinetic form, for times $t < 1/w_p$) proceeds with rapid orientation-flips of the dipoles into their excited low-energy oscillations.[17-18] Subsequently (for times $t > 1/w_p$), each abrupt shock-transition ripples through to the slower many-body resonant flip-flops of the interacting dipole-clusters, dissipating the initial (electrically stored) potential energy to the heat-bath over longer times. Thus, the full relaxation proceeds in two stages; having relatively short "transition" and much longer "adjustment" time-scales respectively.[12, 19]

$CaCu_3Ti_4O_{12}$ is a highly disordered material, where dipoles interact at the mesoscopic scale. To investigate the correlation of dipoles here, we have fitted its dielectric data with the D-H function. The function looks similar but not identical to the Havriliak-Negami (H-N) form.[20-21] It is defined as (WinFit, NovoControl)

$$e^*(w) = e' - ie'' = e_\infty + \frac{e_0 - e_\infty}{F_{01} \cdot (1+iwt)^{n-1}} {}_2F_1\left(1-n, 1-m, 2-n, \frac{1}{1+iwt}\right) \quad (3)$$

Where ${}_2F_1$ is the Gaussian Hyper-Geometric function defined as

$${}_2F_1(a,b,c,z) = \sum_{j=0}^{\infty} \frac{(a)_j (b)_j}{(c)_j} \frac{z^j}{j!},$$

with

$$(a)_j = a(a+1)(a+2).....(a+j-2)(a+j-1).$$

The function $F_{01}$ is defined as

$$F_{01}(n,m) = \frac{\Gamma(2-n)\Gamma(m)}{\Gamma(1+m-n)},$$

$\Gamma$ denoting the Gamma functions. Here we have five fit parameters; $e_0$, $e_\infty$, and $t$ having their usual meanings as in the H-N function, with the indices $n$ and $m$ reflecting the power-law behaviours above and below $w_p$ (eqs.1, 2), and ranging from 0 to1.

We have fitted the D-H function on the dielectric losses and the real permittivity, as shown in fig.1. The fit-parameters $n$ and $m$ exhibit non-trivial $T$-dependences (fig.2) and provide a characteristic temperature for CCTO. The exponent $m$ (slope of the lower-frequency side of the loss-peak in the log-log $e''$ vs. $w$ plot, inset fig.2) is much more sensitive to the temperature, compared to the high-frequency-side slope $n$, showing two distinct behaviours above and below ~100K. At high $T$'s, $m$ tends to its limiting value 1 and $n$ to 0, indicating the tendency towards the (Debyean/symmetric) single-particle dipolar-relaxation of the system, without the correlations. By the same token, the steeper fall of $m$ below ~100K indicates increasing dipolar correlations, also confirmed by the exponent $n$ increasing towards 1, with successively asymmetric loss-peaks. This representation clearly marks 100K as the turning point in the dipolar correlations, where the total dielectric losses too have been observed as the maximum.[22]

To justify the Dissado-Hill form as the most appropriate spectral function representing our data, we compare it with the (force-fitted) Havriliak-Negami[20-21] function; both these computations for our $e''(w)$/100K are displayed in fig.3. Evidently, the D-H function fits well over a broader range of spectrum with physically admissible fitting parameters, contrary to the H-N function where its (asymmetry) index iterates to unrealistic values ($b$ >1). The D-H fitting is also justified from the nature of the relaxation-kinetics (inset, fig.3). Over its entire bandwidth, the Arrot-plot character of the relaxation time $t$ versus $10^3/T$ is the same for both H-N and D-H cases. However, force-fitting $t_{H-N}(T)$ by the Vogel-Fulcher function (V-F, as the corresponding $T$-domain manifestation of a non-Debyean H-N spectrum, $b \neq 1$) produces negative/unrealistic $T_{VF} \cong -58K < 0K$. Rather, a combined exponential-linear fit made on $t_{D-H}(T)$ has been found[22] to be more appropriate, to describe the observed sub-linear Arrot-plot. Thus, from both spectral and kinetic viewpoints, the allied H-N/V-F formulations are confirmed as unphysical/incorrect descriptions of CCTO's particulate dielectric behaviour. On the other hand, the D-H/exp-lin combined framework is amenable to a realistic interpretation.[22]

In conclusion, the nanoscale Ca-Cu site-antisite disorder in $CaCu_3Ti_4O_{12}$ makes it a highly non-linear and electrically heterogeneous material. Here the dielectric response does not follow the single-particle Debyean/KWW patterns, as their corresponding Havriliak-Negami spectral form and the allied $T$-domain Vogel-Fulcher kinetics both confirm out to be physically inadmissible. Rather, the Dissado-Hill formulation signifying dipolar interactions is found to describe the observed dielectric spectra satisfactorily, and parametrically accounts for its behaviour change across the characteristic ~100K. The many-body cooperative two-step relaxation at low temperatures also reconciles with the large drop in its huge permittivity.

**ACKNOWLEDGEMENTS**

The authors thank P. Chaddah and A. Gupta for their support and encouragement.

**FIGURE CAPTIONS**

**Figure 1.** Real and imaginary parts of CCTO-permittivity as a function of frequency at various temperatures. Solid lines shown are the fits according to Dissado-Hill formulation.[11]

**Figure 2.** Dissado-Hill correlation parameters (*m*, *n*) obtained at different temperatures. Parameter *n* describes the power-law behaviour over the frequency range $w > w_p$ and the index *m* applies to the lower range $w < w_p$, as shown in the inset.

**Figure 3.** Comparision of the Dissado-Hill (D-H, solid line) and Havriliak-Negami (H-N, dashed line) fits to the loss-peak at 100K. Lower inset-- consistent with the unphysical asymmetry parameter ***b*** >1 from the H-N force-fit, the correspondingly force-fitted Vogel-Fulcher kinetics [$\ln t_{HN} \sim B/(T-T_{VF})$] too provides unacceptable $T_{VF} \cong -58K$. Upper inset-- Exponential-linear fit[22] to the relaxation time $\ln t_{DH} \sim [aT^{-1} - m\exp(-gT^{-1})]$, obtained from the physically realistic D-H fit.

**Fig.1**

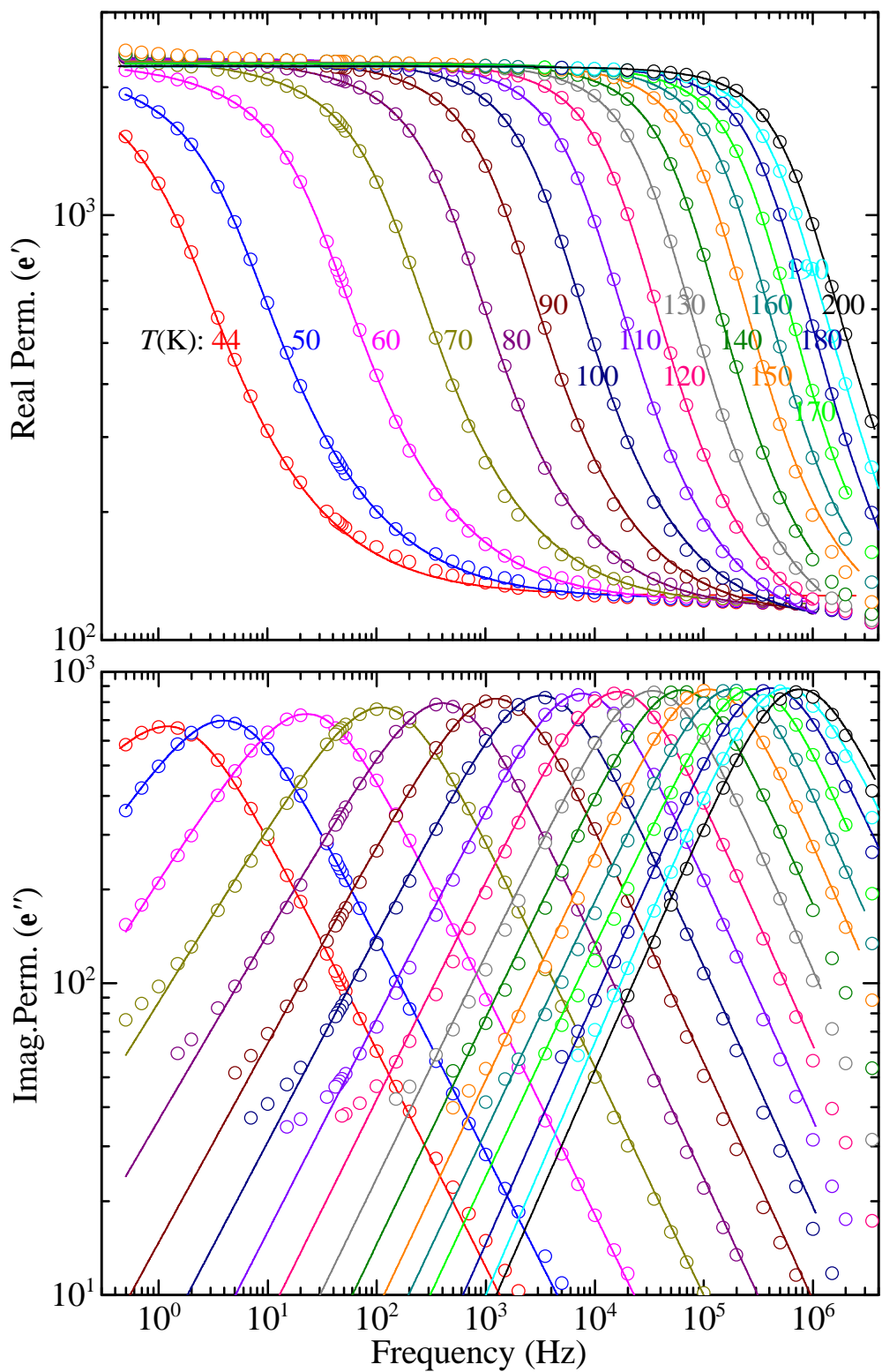

**Fig.2**

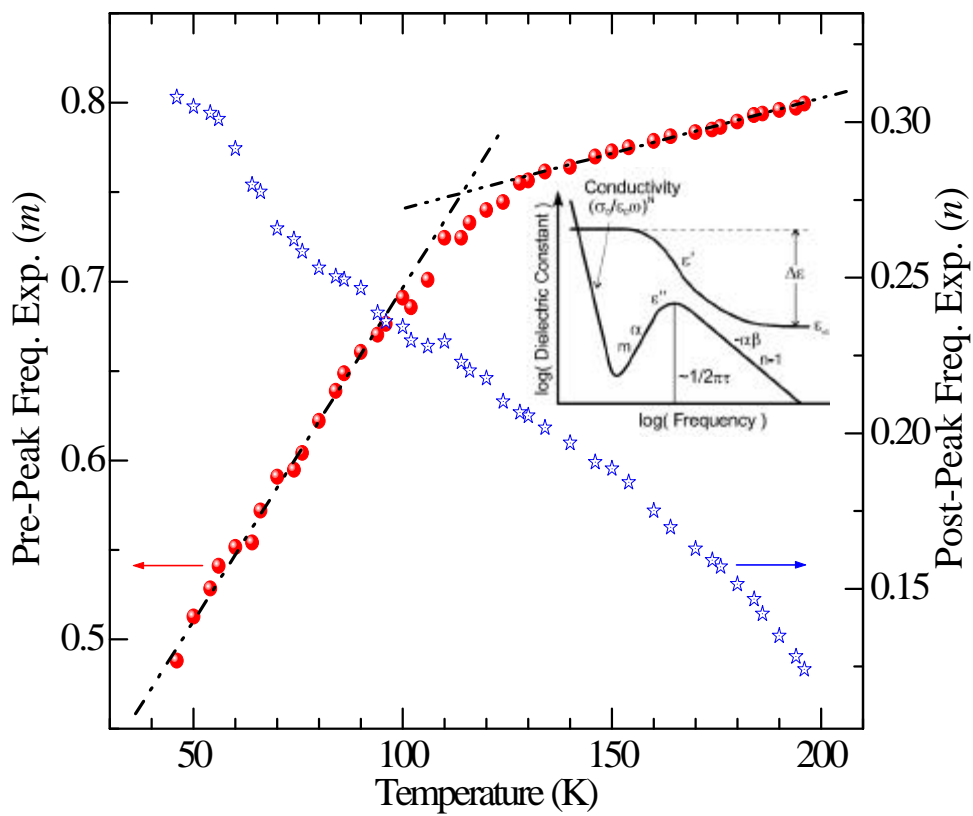

**Fig.3**

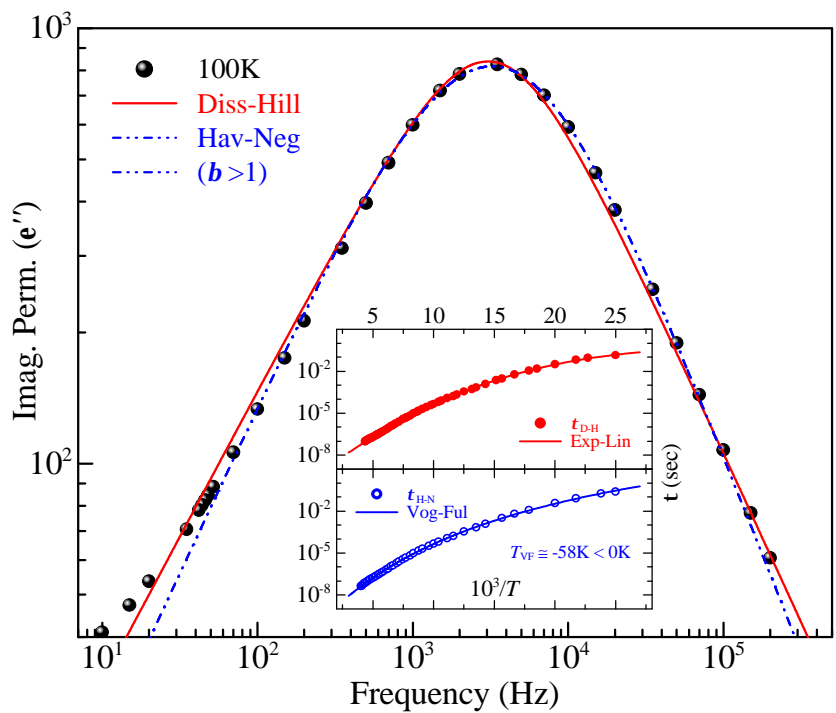